\documentclass[vecphys]{svmult}

\usepackage{makeidx}         % allows index generation
\usepackage{graphicx}        % standard LaTeX graphics tool
\usepackage{multicol}        % used for the two-column index
\usepackage[bottom]{footmisc}% places footnotes at page bottom
\makeindex             % used for the subject index
\begin{document}

\title*{Origin theories for the eccentricities of extrasolar planets}
\titlerunning{Eccentric extrasolar planets}
% your contribution title if the original one is too long
\author{Fathi Namouni\inst{1}}
% Use \authorrunning{Short Title} for an abbreviated version of
% your contribution title if the original one is too long
\institute{CNRS, Observatoire de la C\^ote d'Azur, BP 4229, 06304 Nice, France
  \texttt{namouni@obs-nice.fr}}
%
% Use the package "url.sty" to avoid
% problems with special characters
% used in your e-mail or web address
%
\maketitle

\section{Introduction}
Planetary science has known a revolution in the year 1992 when Wolszczan and
Frail discovered the first extrasolar planets around a pulsar \cite{c1}. This
discovery was soon followed in 1995 by the first Jupiter-like planet around
the sun-like star 51 Pegasi \cite{c2}. As of this writing we have entered the
stage of statistics as more than 200 extrasolar planets are known to orbit
sun-like stars. The revolution that was triggered by these discoveries changed
the paradigm of what we think of as a planetary system and what we believe to
be the scenarios that led up to the formation of planets. The planet around 51
Pegasi is the prototype of what has become to be known as hot Jupiters:
planets with masses comparable to Jupiter's but with periods of a few days and
orbits smaller than the tenth of Mercury's distance to the sun.  Planetary
orbits also seem to be eccentric. This observation is not surprising since the
solar system planets have eccentric orbits except that half the extrasolar
planets have eccentricities larger than 0.3 which is much more significant
than Jupiter's 0.05. Such large eccentricities are reminiscent of the small
body populations of the solar system that got stirred up by their
gravitational interactions with the larger planets. In this respect,
extrasolar planetary eccentricities are more unusual, witness the median
eccentricity of 0.13 of main belt asteroids larger than 50\,km. The planetary
revolution did not stop at orbital radii and eccentricities: 10\% of known
planetary systems belong to binary star systems and only one planet so far is
in a triple stellar system.  Multiple planets around a single star make up
about 10\% of known planets.

In essence, the planetary revolution heralded the coming of a new planetary
principle: orbital diversity is a rule of planetary formation. Diversity here
is not meant to imply subtle changes but drastic ones with respect to the
aspect of the solar system. This state of affairs has prompted a serious
revision of the theories of planetary formation: hot Jupiters with few-day
orbits could not have formed in situ.  Instead they have formed outside a few
AU for a sun-like star where it is possible for ices to condense and for the
planets to capture large amount of gas from the protoplanetary disk.  Only
after they formed, did they travel all the way to meet their current orbits.
It is interesting to note that the concept of radial migration was already
known in the contexts of accretion disks \cite{c3} in binary star systems and
of planetary rings \cite{c4}. Only before 1995, one could not plausibly
contemplate the prospect of suggesting the existence of massive planets that
traveled all the way from Jupiter's current location just to stop on a close
orbit with a few-day period. The basic aspects of the process of planetary
migration through the tidal interaction of a planet with the gaseous
protoplanetary disk are now well understood \cite{b25} yet an important
challenge remains: what stops planetary migration towards the star?  The
leading contender for stopping planetary migration is the planet's interaction
with the stellar magnetosphere but a definitive quantitative description is
still lacking.

Extrasolar planetary eccentricities have equally resulted in a drastic change
of perception: it is often heard that it is not the extrasolar planets that
are eccentric, rather it is the solar system that lacks eccentricity. This
perception is encouraged by the availability of some simple instabilities that
one can set up in a many-body gravitational system to simulate the generation
of the wild orbits of extrasolar planets. Upon close examination such
instabilities as well as other eccentricity scenarios do not tell the whole
story of how extrasolar planets become eccentric. In fact, just as the
features of the planetary migration process yield constraints on the planetary
formation scenarios, so do the various theories of the eccentricity
excitation.

It is the aim of this chapter to review the various processes of the origin of
extrasolar planets' eccentricities in the context of planetary formation.  We
start by reviewing the properties of extrasolar planetary orbits in section 2.
Section 3 contains a commentary on the various known theories of eccentricity
excitation. Section 4 specializes in a recent addition to the eccentricity
theories based on an relationship between the planets and the stellar jet that
is powered by the protoplanetary accretion disk. The final section 5 discusses
how the eccentricity origin problem may contribute further to the theory of
planet formation.

\section{Eccentricity observations}
\begin{figure}
\centering
% Use the relevant command for your figure-insertion program
% to insert the figure file.
% For example, with the option graphics use
\includegraphics[height=5.5cm]{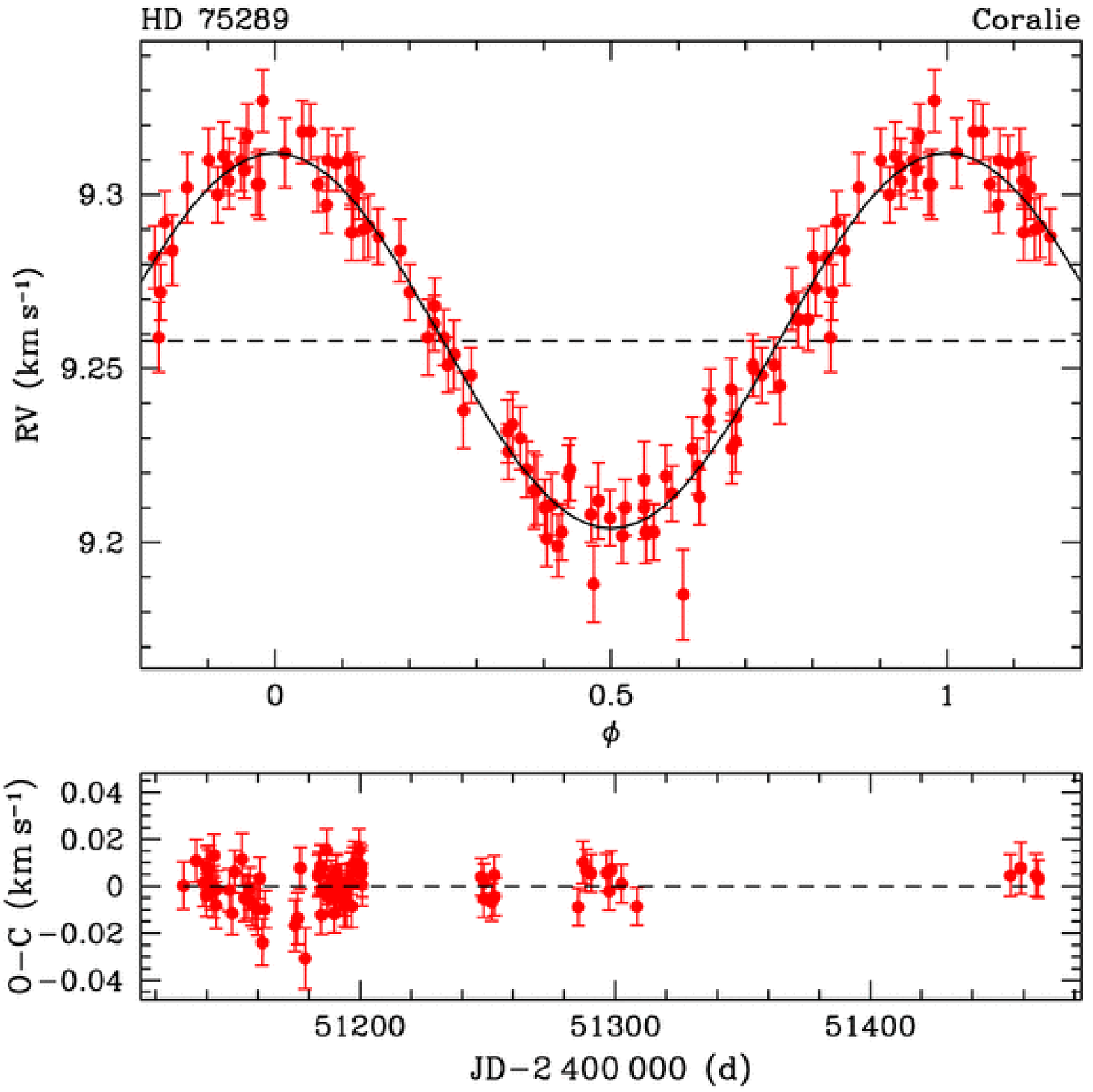}
\includegraphics[height=5.5cm]{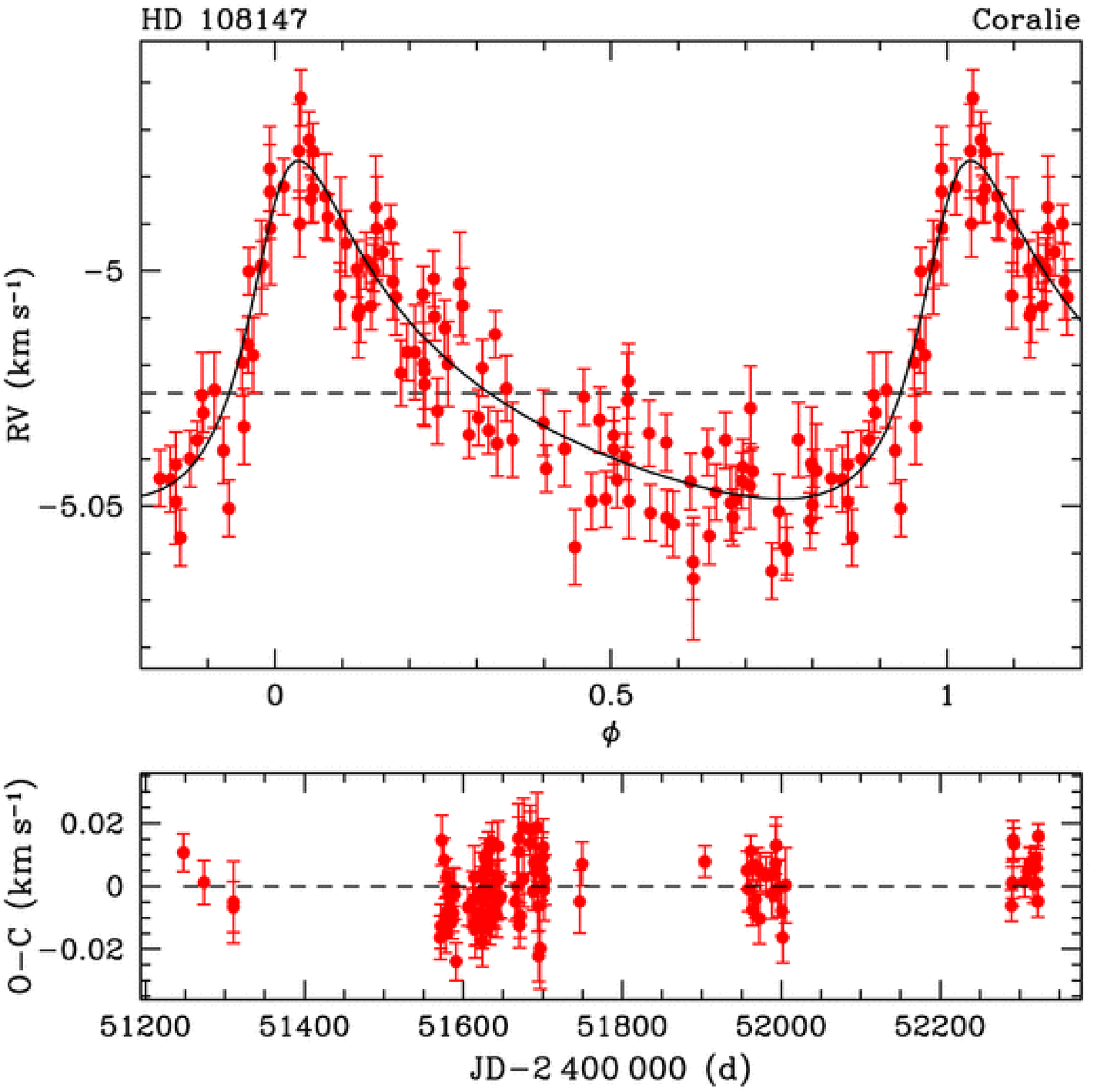}
%
% If not, use
%\picplace{5cm}{2cm} % Give the correct figure height and width in cm
%
\caption{
  Reflex velocity of the stars HD 75289 (left) \cite{c5} and HD 108147 (right)
  \cite{c6}. The planet around HD 75289 has a circular orbit while that around
  HD 108147 has an orbital eccentricity of 0.5. Pictures taken from the Geneva
  Extrasolar Planet Search {\it http:$/\!\!/$obswww.unige.ch$/$\~ \ 
    $\!\!u$dry$/$planet$/$planet.html} }
\label{fig1}       % Give a unique label
\end{figure}

Extrasolar planets are detected with various observational techniques
\cite{b20}. The Doppler analysis of the reflex velocity of the host star is by
far the most successful technique to date. It is also the technique that has
uncovered the large eccentricities of extrasolar planets. If a planet has a
circular orbit, the analyzed stellar spectrum yields a sinusoidal oscillation
of the stellar reflex motion. If the planet is on an eccentric orbit, the
reflex motion as a function of time becomes distorted with respect to a pure
sine reflecting the unequal times the star spends in different locations along
its orbit around the center of mass of the star-planet system (Figure
\ref{fig1}).  The discovery of large eccentricity orbits by the Doppler reflex
velocity method is due to its ability to detect planets on wider orbits in
contrast to that to the transit method. Planets on many-day periods have
usually undergone tidal circularization by the host star.
%%%%%%%%%%%%%
%%%%% FIG 1

%%%%%%%%%%%%%

The statistical analysis of extrasolar planet eccentricities reveals very few
clues as to the origin of the elongated orbits. For the known sample of 196
planets discovered by the reflex velocity and transit techniques, the median
eccentricity is at 0.21 if all planets are counted and at 0.28 if planets with
periods smaller than 5 days are excluded because their circular orbits simply
reflect tidal circularization. The prevalence of such large eccentricities and
the large typical mass of the detected planets (comparable to Jupiter's) has
encouraged the comparison of the extrasolar planetary systems to binary star
systems.  Depending on the methods used, similarities in the eccentricity
distribution of both populations can be found \cite{b22} or not \cite{c7}.
What is agreed upon is that there is no correlation between the size of the
orbits and their eccentricities in each population, and no striking
resemblance of the scatter of both populations in an orbital size versus
eccentricity plane. The size of the orbit usually refers to either the
semi-major axis or the pericentre radius. The latter is used to account for
those orbits that have not yet had enough time to be circularized --as the
pericentre distance is conserved under stellar tides. Finally, eccentricities
show a vague correlation with the planetary masses with heavier planets
enjoying larger eccentricities.

\section{Eccentricity origin theories}
Seven known explanations have been put forward to account for the large
eccentricities of extrasolar planets. They are: (1) planet-planet scattering,
(2) planet-protoplanetary disk interaction, (3) Kozai's secular cycles, (4)
excitation through radial migration into a mean motion resonance, (5) Stellar
encounters, (6) stellar-like N-body relaxation, and (7) excitation through
stellar jet acceleration. In the following, we comment on these possibilities by
discussing their instability types, characteristic timescales, their epoch of
applicability as well as their advantages and drawbacks.

\subsection{Planet-planet scattering}
Planet-planet scattering is a simple process to generate eccentric orbits in
an N-body gravitational system. If a system of two or more planets on planar
circular orbits find themselves ``initially'' closer than is permitted by
Chirikov's criterion for the overlapping of mean motion resonances \cite{c12},
the planets scatter off one another leading to a system with more stable
albeit eccentric configurations. Depending on the number, masses and
``initial'' spacings of the planets, the instability timescale varies between
$10^3$ to $10^7$ years \cite{c13,c14,c15,c16,c17}. The epoch that is referred
to by the adjective ``initial'' is that of the disappearance of the agent or
the conditions that kept the planets from scattering off one another in the
first place. This epoch is customarily associated with a significant dispersal
of the parent gaseous protoplanetary disk. As well shall point out in the next
section, planet-disk interaction is known to primarily erase orbital
eccentricities.  An additional condition for planet-planet scattering to be
operational is the absence of a significant population of smaller bodies such
as the primordial asteroid belt. Depending on the mass spectrum in the
planetary system, the smaller populations are able to limit the growth of the
planetary eccentricities through dynamical friction \cite{c8}. This at least
how it is believed that the terrestrial planets in the solar system did not
acquire large eccentricities \cite{c11,c10}. Numerical works that tackle the
extrasolar eccentricity problem using planet-planet scattering do not consider
the effect of leftover small-body populations after the gaseous disk has
dispersed. Planets are set up at a few Hill radii from one another and initial
conditions are sampled to reproduce the eccentricity of certain observed
systems. The general excitation trend of planet-planet scattering leads to
larger eccentricities than the ones observed. What may prove to be a serious
problem for planet-planet scattering is the eccentricity distribution obtained
in multiple systems that contain Jupiter-mass planets as well as Earth-mass
planets. The conservation of angular momentum in this case will force the much
smaller planets to have much larger eccentricities than the Jovian planets.

\subsection{Planet-disk interaction}
A planet embedded in a gaseous disk excites sound waves at the locations of
its mean motion resonances within the disk akin to the gravity waves
excited by Saturn's satellites in its ring system. The density enhancements at
the mean motion resonances act back on the planet resulting in gravitational
torque. Two types of resonances contribute to this torque: (1) corotation
resonances that primarily affect the semi-major axis and tend to damp any
acquired eccentricity and (2) Lindblad resonances that primarily affect the
eccentricity and tend to increase it \cite{c4-0,c4}. The torque contribution
of the former is larger than the latter's by about 5\%. At first sight,
planet-disk interaction damps the eccentricity on timescale that depends
strongly on the disk's thickness and less strongly on the disk's mass density
and the planet's mass \cite{b25b,c4-2,c4-21}. The torques originating from
higher order resonances as well as those pertaining to the relative
inclination of the planet and the disk do not change the outcome significantly
\cite{b29,c10b}.  Only if the corotation torque saturates, can the Lindblad
resonances increase the eccentricity \cite{b29,c4-1}. The conditions under
which saturation arises are difficult to quantify explaining why an eccentricity
increase due to a disk-planet interaction has never been observed in numerical
simulations although this might be due to numerical artefacts \cite{c4-3}.

\subsection{Secular Kozai cycles}
In his study of asteroids perturbed by Jupiter on high eccentricity and
inclination orbits, Kozai \cite{b8-0} showed that the averaging of the
interaction potential over the mean motion without expanding the force
amplitude in terms of eccentricity and inclination leads to new types of
secular resonances. The conservation of the vertical component of angular
momentum (vertical refers to the direction of Jupiter's orbital normal) shows
that when the orbital eccentricity increases, the inclination decreases. In
particular, if objects are set up on inclined but circular orbits, large
eccentricities can be achieved as the inclination decreases in its motion
around the secular resonance cycle. The application of the Kozai cycle to the
eccentricities of extrasolar planets assumes that there is a binary star on a
not-too-far inclined orbit that perturbs the planet that formed in a circular
orbit in a timescale shorter than the Kozai libration cycle.  In essence, the
secular Kozai cycle idea transforms the eccentricity problem into an
inclination problem. In this sense, the observed planets do not possess a
proper eccentricity but one that is forced by the stellar binary and that will
always oscillate between its original value, zero, and a maximum value
depending on the planetary-binary semi-major axis ratio, the binary's mass and
its orbital inclination with respect to the plane on which the planet
initially formed.  When applied to specific binary star systems with one
planet, the Kozai mechanism works fine and helps characterize the orbit and
mass of the secondary star required to excite eccentricity
\cite{b8,b8-1,b8-2}.  Statistically, Kozai based excitation of one-planet
binary systems tend to yield larger eccentricities than observed \cite{b8-3}.
As T Tauri stars form in multiple systems, it is not unreasonable to try and
apply the Kozai mechanism to the whole sample of observed extrasolar planets.
The problem is that the Kozai cycle is usually destroyed by mutual
gravitational interactions. The addition of more planets to the one-planet
binary system, forces the precession of the planets' pericentres. If the
planets are of comparable mass as it is observed in multiplanet systems, the
Kozai cycle is lost.

\subsection{Mean-motion resonances}
The role of mean motion resonances in exciting orbital eccentricity has its
roots in the study of the orbital evolution of Jupiter's and Saturn's regular
satellites under planetary and satellite tides \cite{b3-1}. These satellite
systems are known to be in or to have crossed mean motion resonances thereby
acquiring forced eccentricities. The combined modeling of the orbital
evolution, capture into resonance and the tidal interaction lends valuable
bounds on the dissipation factors of Jupiter, Saturn and their satellites.
Extrasolar planets form in a gaseous disk that does not dissipate after they
acquired most of their masses or else hot Jupiters would not exist.
Planet-disk interaction naturally gives rise to orbital migration with
different planets in the same system migrating at different rates. This
differential migration makes planets in the same system encounter mean motion
resonances.  Capture into resonance may occur depending on whether the
migration is convergent or divergent (for instance if the outer planet is
moving faster or slower than the inner planet). Convergent migration leads to
capture into resonance. The subsequent common migration of a planetary pair in
resonance pumps up the eccentricities on the migration timescale
\cite{b11,b3-2}. Divergent migration does not lead to resonance capture,
instead eccentricity jumps are acquired at resonance passage \cite{b3,b3-3}.
While convergent migration is certainly the way the known resonant multiple
systems have acquired their eccentricities, this excitation method involves a
mystery that may shed light on how to halt planet migration in a gas disk. The
mystery consists of the observation that convergent migration is far too
efficient in exciting eccentricities to the point where in many systems, when
capture occurs, migration must stop quickly thereafter or else eccentricities
are pumped up to much larger values than those observed. As it is implausible
to invoke the dispersal of the gas disk, planetary migration may become
ineffective because of the nonlinear response of the gas disk to the planet
pair. It is interesting to note that when capture occurs, the planetary
relative inclinations may be excited as unlike planetary satellites that orbit
Jupiter and Saturn, the central potential is keplerian. Consequently, for the
same order, eccentricity and inclination resonances are close (but not
coincident as the gas disk modifies the pericentre and node precession rates).
Planet-disk interaction is not well understood for large eccentricity planets
and off-plane (inclined with respect to the disk) orbits. The often used
formulas for eccentricity damping from the disk torques have not been verified
for eccentric and inclined planets. Divergent migration has the advantage of
being applicable to the wider non-resonant multiplanet systems.  For a
planetary pair, divergent migration requires that the inner planet migrates
faster than the outer one. Gap driven migration (also known as type II) is
favorable to such a condition as the migration rate is the viscous timescale
of the disk. Divergent migration may take place because viscosity is likely to
be a decreasing function of the distance to the star. If however the part of
the disk that is located between the two planets is dispersed as when the two
planetary gaps merge, the direction and rate of migration may be altered
significantly \cite{c4-31}.

\subsection{Stellar encounters}
Stellar encounters are common events in star clusters. A planetary system that
encounters a star will have its planets feel a tidal force that elongates
their orbits . For inner planets that orbit close to the host star, the
excitation which lasts for about 1000 years will occur on a secular timescale.
Outer planets if they exist will feel a localized impulse somewhere in their
orbits.  Typical encounter frequencies are of one in $5\times 10^9$ years
while typical encounter parameters are a few hundred AU. Unless planets are
way outside the classical planetary region (inside 30 AU), excitation is not
efficient \cite{b27}. To reverse this conclusion and account for the
eccentricities of inner planets, the system must contain several planets with
increasing distance and mass from the star in order propagate the stellar tug
felt by the outermost planet down to the innermost ones \cite{b27}.

\subsection{Stellar-like relaxation}
The qualitative similarity of the eccentricities of extrasolar planets and
stellar binaries suggests that planets may form through similar processes as
those of multiple stellar systems. If planets formed by gravitational
instability, the formation time is so short that the planets find themselves
confined to a smaller space than their orbital stability permits. The
relaxation of such systems leads to some planetary ejections and many large
eccentricity orbits \cite{b19}. The applicability of this scenario is limited
by two facts: first, the minimum planet mass the gravitational instability
allows is a few Jupiter masses. This means that stellar-like relaxation does
not work for planets with masses comparable to or smaller than Jupiter's.
Second, if a two-phase formation where small planets form through rocky core
accretion and the larger ones through gravitational instability \cite{b23},
then it is likely that the relaxation of the larger planets destroys the
smaller planets. This is because the gravitational instability timescales are
usually smaller than the planetesimal accumulation timescales. In fact, if
large mass planets form through gravitational instability, they are likely to
inhibit planetesimal accumulation by clearing the inner disk before planetary
embryos are born.

\section{Jet-induced excitation}

Stellar jets enter the eccentricity excitation problem because of their
ubiquity and simplicity \cite{c10-1}. The story of how this works is as
follows: although there is disagreement on whether there is a statistically
significant resemblance between the eccentricity distributions of extrasolar
planets and stellar binary systems, the qualitative similarity is beyond
doubt. Those who wish for the similarity to be quantitative, would like to
affirm the view that planets are the lower end of the outcome of star
formation. This question has been settled observationally in 2005 with two
observations: the first is a hot Saturn with a giant rocky core discovered by
combining the Doppler reflex velocity method with transit photometry
\cite{d3}. The second is the imaging of the first planetary candidate which
because of bias due to contrast and resolution happens to be a warm distant
companion orbiting a young brown dwarf \cite{d4}. This proves that planets do
not need large rocky cores and may form by gravitational instability. Exit the
link between how planets form and their eccentricities.

If planets do not form like binary stars, perhaps they undergo similar
excitation processes that lend them similarly elongated orbits. In view of the
different physical environments where planets and stars form, the simplest
possible excitation process may depend weakly or not at all on the local
dynamics of the stellar or planetary companion. Mathematically, this amounts
to saying that the acceleration imparted by the process is independent of
position and velocity.

Simplicity therefore dictates that the process imparts a constant acceleration
that operates during a finite time window. Simplicity also comes with two
added advantages: we can already know the excitation time scale and the
minimal acceleration amplitude. Dimensional analysis shows that the excitation
timescale has to be proportional to $v/A$ where $v$ is the keplerian velocity
of the companion around the main star. Further, if the acceleration is to
achieve its purpose within the lifetime of the system, $v/A$ must be smaller
than about $ 10^9$ years. This tells us that the acceleration $A> 3\times
10^{-16}(v/10\,\mbox{km\,s}^{-1})\ \mbox{km\,s}^{-2}$.

The process lacks one more attribute: direction. If the acceleration is
independent of the formation processes, its direction cannot depend on
anything related to the planetary companion such as its orbital plane or the
direction from the star to the companion. In an inertial frame related to the
planetary or stellar system, we are not left with much choice but the star's
rotation axis.

To sum up, what we are looking for is a process that appears everywhere where
planet and star formation takes place, acts like a rocket (i.e. with an
acceleration that does not depend on the position and velocity of the system)
and whose direction is related to the star's rotation axis. The answer is
stellar jets \cite{b52,b53}. 

Do planets exist when jets are active?  The answer is quite likely. Known hot
Jupiters have moved close to their host stars because of their interaction
with the gas disk. So we know the gas disk was present and had viscosity well
after planets finished forming. The gas accreting on the star because of
viscosity is the main ingredient along with the magnetic field that threads it
needed to launch stellar jets and disk winds.  It would therefore be an
interesting coincidence that jets shut off when planets appear in the gas disk
a few AU away from the star well outside the jet launching region.

Are there any observational hints that jet-sustaining disks contain planets?
The only possible hint so far is the observation of variable brightness
asymmetries in some jet-sustaining disks \cite{hh1,hh2,hh3}. The variability
timescales of a few days to a few years are so small that they imply either a
peculiar stellar activity in the form of single hot spots or the presence of
distortions in the disk at the location where the orbital period matches the
variablity timescale. The first option requires a complex stellar magnetic
field that differentiates strongly between the two stellar poles. The second
option may be caused distortions in the disk whose origin could be the
presence of embedded compact objects.

Do jets have enough strength to build eccentricity? Inferred mass loss rates
for known young T Tauri stars lie in the range $\sim 10^{-8}M_\odot\,
\mbox{\rm year}^{-1}$ to $10^{-10}M_\odot\, \mbox{\rm year}^{-1}$ and may be
two orders of magnitude larger depending on the way the rate is measured from
the luminosity of forbidden lines \cite{b46,b48,b47}. The jet also needs to be
asymmetric with respect to the star's equator plane or else there would be no
acceleration. Interestingly, a growing number of bipolar jets from young stars
\cite{b50,b50-1,b49,b49-1} are known to be asymmetric as the velocities of the
jet and counterjet differ by about a factor of 2. Mass loss processes in young
stars therefore yield accelerations:
\begin{equation}
A \sim  10^{-13}\, 
\left(\frac{\dot M}{10^{-8} M_\odot\, \mbox{\rm  year}^{-1}}\right)\, 
\left(\frac{v_e}{300 \,\mbox{\rm km\,s}^{-1}}\right)\,
\left(\frac{M_\odot}{M}\right) \mbox{\rm km\,s}^{-2}. 
\end{equation}
where $M$ is the stellar mass and $v_e$ is the outflow's high velocity
component. As jets are time-variable processes, the above estimate is only
indicative of the epochs at which the rates and velocities are measured. In
this sense, it is closer to being a lower bound on what accelerations really
are over the jet's lifetime. What is clear is that asymmetric jet acceleration
is larger than the minimum amplitude of $10^{-16} (v/10\,\mbox{km\,s}^{-1})\ 
\mbox{\rm km\,s}^{-2}$.

For how long can a jet operate? Jet-induced acceleration is technically no
different from attaching a rocket to the star and accelerating it very slowly
with respect to the outer part of the disk and the planets. As a result, the
star acquires a residual velocity that must be smaller than its orbital
velocity in the Galaxy or else the star is ejected.  In fact, there is an even
a stronger constraint on the residual velocity from the velocity dispersion in
the Galaxy. Stars do not have exactly circular orbits in the Galaxy. Their
motion is slightly distorted or eccentric and such eccentricity is measured as
a departure of the galactic orbital velocity from that of circular motion.
This velocity dispersion is known for various stellar populations and is of
order a few tens of kilometers per second. Since the residual velocity
imparted by the jet, $V$ is given by the product $A \tau$ where $\tau$ is the
duration of acceleration, imposing that $V<\left<v_g\right>$ where
$\left<v_g\right>$ is the velocity dispersion in the Galaxy yields:
\begin{equation}
\tau \leq 
10^5 
 \ \frac{3\times
10^{-12}\,{\rm km\,s}^{-2}}{A}\ \frac{\left<v_g\right>}{10\, {\rm km\, s}^{-1}} \, {\rm years}.
\end{equation}
This timescale is shorter than the disk's lifetime. In practice, we shall see
that shorter times are needed.

Further excitation properties can be deduced by analyzing the effect of the
combined jet-induced acceleration and the star's gravitational attraction. As
the star's pull decreases with distance, there is a specific location where
the latter matches the jet-induced acceleration (that is independent of
position and velocity). Outside this radius, the star's pull is weak and
orbits escape its gravity. This reveals an interesting feature of jet-induced
acceleration: stellar jets are responsible for the outer truncation of
circumstellar disks. It is clear that in the interior vicinity of the
truncation radius, the orbital perturbations are large as the excitation time
becomes comparable to the orbital period. In this region, the keplerian orbits
are subject to a sudden excitation; not only the eccentricities are excited
but the semi-major axes are also affected leading to inward or outward
migration.  Well inside the truncation radius, the excitation time is much
larger than the orbital period. In this region, eccentricity builds up slowly
over a large number of revolutions of the planet around the star and the mean
orbital radius remains constant on average.  Excitation in this region occurs
on secular timescales.  Planetary companions mostly fall inside the secular
region as they are far inside the truncation radius which is more or less the
size of the protoplanetary disk.

\subsection{Secular jet-induced excitation}

\begin{figure}
\centering
\includegraphics[height=5cm]{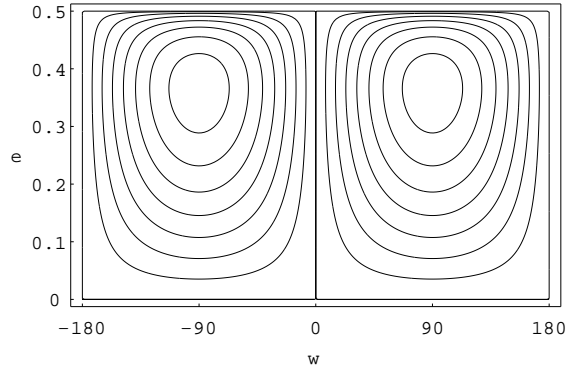}
\caption{Contour plots of the acceleration potential (\ref{secpot2}) in the
         eccentricity $e$ and argument of pericentre $\omega(^\circ)$
         plane. The direction of acceleration makes an angle $I_0=30^\circ$
         with respect to the companion's angular momentum vector. The time
         evolution of the two orbits ($e=0$, $\omega=0$) and ($e=0.3$,
         $\omega=90^\circ$) is shown in Figure (\ref{fig3}).}
\label{fig2}
\end{figure}

In the secular region where the excitation time is larger than the orbital
period, the dynamics of excitation can be simplified by averaging the
acceleration over the orbital period of the companion. For a constant
acceleration, the interaction potential is simply $R={\vec A}\cdot {\vec x}$
where ${\vec x}$ is the position vector. Averaging the interaction potential
amounts to averaging the position vector of a pure keplerian motion. A simple
calculation shows that $\left<{\vec x} \right>=-3 a e\, {\vec x}(f=0)/2r$
where $f$, $a$, $e$ and $r$ are the true anomaly, the semi-major axis, the
eccentricity and radius of the keplerian orbit. The direction of $\vec x$ at
pericentre is that of the eccentricity vector ${\vec e}={\vec v}\times {\vec
  h}/G(m+M)-{\vec x}/|{\vec x}|$ where ${\vec v}$ is the velocity vector of
the companion, $G$ is the gravitational constant, $m$ and $M$ are the masses
of the companion and the host star and ${\vec h}={\vec x}\times {\vec v}$ is
the specific angular momentum. This enables us to write the secular potential
as:

\begin{equation}
\left<R\right> =-\frac{3}{2}
 a \, {\vec A}\cdot {\vec e}=-
\frac{3}{2}
A\, a e \sin(\varpi-\Omega) \sin I,
\label{secpot}
\end{equation}
where in the last equality, the $z$--direction of the reference frame is chosen
along along $\vec A$ and $\varpi$, $ \Omega$, $I$, are the longitude of
pericentre, longitude of ascending node and the inclination of the orbit. To
simplify the excitation problem further, we use the conservation of the
component of angular momentum $\vec h$ along the direction of acceleration as
$\vec A \cdot \dot{\vec h}=\vec A \cdot(\vec x \times \vec A)=0$. In the
reference frame where $\vec A$ is along the $z$--direction, the conservation
of angular momentum yields $(1-e^2)^{1/2}\cos I=\cos I_0$ where $I_0$ is the
initial inclination of the keplerian orbit with respect to the jet-induced
acceleration. This relation enables us to eliminate the inclination variable
in $\left<R\right>$ and reduce the problem to an integrable, one dimensional
system with:

\begin{equation}
\left< R\right>= -\frac{3}{2}
A\, a  \,
     \sqrt{\frac{\sin^2I_0-e^2}{1-e^2}}\, e \sin \omega. \label{secpot2}
\end{equation}
where $\omega=\varpi-\Omega$ is the argument of pericentre. The time evolution
of eccentricity and argument of pericentre is obtained from:
\begin{equation}
\dot e=-\frac{\sqrt{1-e^2}}{na^2\,e}\ \frac{\partial\left< R\right>}{\partial
     \omega}, \ \ \ \dot \omega=\frac{\sqrt{1-e^2}}{na^2\,e}\
     \frac{\partial\left< R\right>}{\partial e}, \label{edotomdot}
\end{equation} 
where $n=\sqrt{G(M+m)/a^3}$ is the companion's mean motion. In this
one-dimensional system, $e$ and $\omega$ follow curves of constant $\left<
  R\right>$ shown in Figure (\ref{fig2}). There are equilibria at $\omega=\pm
90^\circ$ and $e=\sqrt{2} \sin (I_0/2)$ corresponding to
$I=\cos^{-1}(\sqrt{\cos I_0})$. The maximum value of $e$ is $\sin I_0$ and
corresponds to the cycle of initially circular orbits. For these orbits,
$\left< R\right>=0$ throughout their cycle implying that the orbits
orientation can take only one value $\omega=0$ modulo $180^\circ$. 

%%%%%%%%%%%
%%%  FIG  2
%%%%%%%%%%%

\begin{figure}\centering
\includegraphics[height=3.3cm]{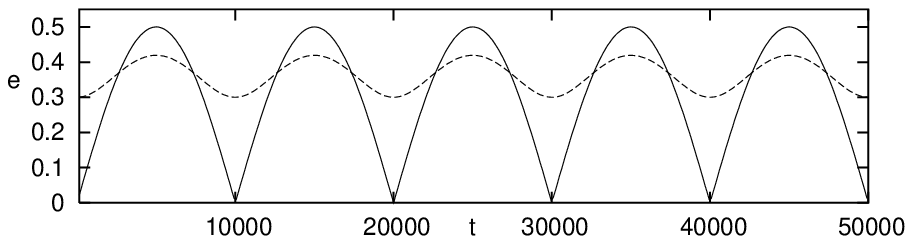}\\
\includegraphics[height=3.3cm]{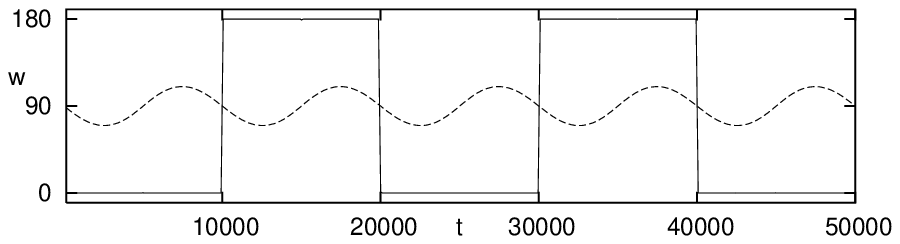}\\
\includegraphics[height=3.3cm]{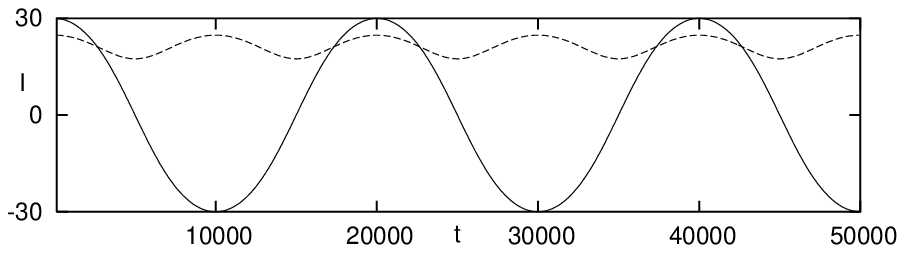}
\caption{Time evolution under a conservative acceleration. The eccentricity
  $e$, argument of pericentre $\omega(^\circ)$ and inclination $I(^\circ)$ are
  shown for an initially circular orbit $e=0$ (solid) and an orbit librating
  about the secular resonance $\omega=90^\circ$ with an initial eccentricity
  $e=0.3$ (dashed).  The semi-major axis is identical for both orbits and is
  set to unity.  The acceleration corresponds to a period of $10^4$ years at
  1\,AU. The plots were obtained by the numerical integration of the full
  equations of motion.}
\label{fig3}
\end{figure}

For time-dependent accelerations and provided that the variation timescale is
longer than the orbital period, the eccentricity evolution is given by:
\begin{equation}
\dot e=\frac{3A(t) \epsilon}{2na}\ 
\,
     \sqrt{\sin^2I_0-e^2}, \label{edot}
\end{equation}
where $\epsilon$ is the sign of $\cos\omega$ which is set by the requirement
that $e\geq0$. The solution of (\ref{edot}) can be found exactly as:
\begin{equation}
e(T)= \left|\sin\left[ \frac{3}{2na}\int_{-\infty}^{T}\,
 A(t)\,{\rm d}t\right]\ \sin I_0 \right|.
\label{eoft}
\end{equation}
The inclination is obtained from $\cos I= \cos I_0/\sqrt{1-e(T)^2}$.  For
strictly constant accelerations (infinite time window), $A(t)=A_0$ and $e$
oscillates between $0$ and $\sin I_0$ at the excitation frequency:
\begin{equation}
n_A=\frac{3\left|A_0\right|}{na}\,
  . \label{nsec}
\end{equation}
Examples of such oscillations that were obtained from the direct integration
of the full equations of motion are shown in Figure (\ref{fig3}). The good
agreement between the secular solution and the results of the numerical
integration comes from the fact that $A$ is independent of position and
velocity.

%%%%%%%%%%%
%%%  FIG  3

\begin{figure}\centering
\includegraphics[height=3.3cm]{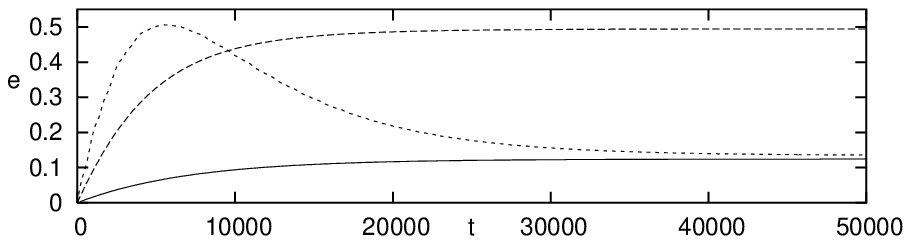}
\caption{Eccentricity excitation by time-dependent constant-direction
  accelerations. The equations of motion are integrated numerically with an
  acceleration $A(t)=A_0H(t)\exp-t/\tau$ where $H(t)$ is the Heaviside unit
  step function and $A_0=2.21\times 10^{-11}$\,km\,s$^{-2}$.  The oscillation
  period at 1\,AU is $1.11\times 10^5$ years.  The timescale $\tau=7200$ is
  chosen so that $V=$5\,km\,s$^{-1}$.  The curves correspond the semi-major
  axes: 1\,AU (solid), 32\,AU (dashed) and 128\,AU (dotted).}
\label{fig4}
\end{figure}

%%%%%%%%%%%
To optimize the excitation of a finite eccentricity from an initially circular
state, the duration of acceleration needs to be smaller than half the
oscillation period: $\tau<\pi na/3|A_0|$. Examples of eccentricity excitation
at three different semi-major axes (i.e. three different excitation
frequencies) are shown in Figure \ref{fig4}. The dependence of the excitation
amplitude on the ratio of the duration to the excitation time is also seen in
the same Figure.  Note that because eccentricity excitation in the secular
region is a slow process compared to the orbital time, the convolution of the
dynamics under the constant acceleration $A$ with a finite time window has the
effect of shutting off the excitation at some eccentricity value depending on
the duration.
%%%%%%%%%%%
%%%  FIG  4  

%%%%%%%%%%

The secular excitation through jet-induced acceleration is therefore able to
make $e$ reach $\sin I_0$ and is largest if the initial orbital plane contains
the direction of acceleration ($I_0=90^\circ$).  As $\left< R\right>=0$ for
initially circular orbits, $\omega$ and $\Omega$, remain at zero. This forcing
of the pericentre to be perpendicular to the direction of acceleration favors
apsidal alignment in multiplanet system. 

If the jet's inclination with respect to the companion's orbital plane is
small $I_0\ll 1$, the maximum eccentricity will be negligible. The $\sin I_0$
limitation is problematic because it is not reasonable to expect stellar jets
to be highly inclined with respect to the gas disk where the companions form.
Fortunately for the jet-induced excitation theory, there is a natural way out.
Jets are known to precess over timescales from $10^2$ to $10^4$ years
\cite{b55,b54,b51}. The origin of such precession is not known as we lack
resolution to probe inside the jet launching region. It is possible that
precession is caused by a warp in the disk's plane resulting from interactions
with stellar companions as T Tauri stars are known to form in multiple
systems. Precession is attractive because it offers the possibility of
resonance if the excitation frequency $n_A$ matches the jet precession
frequency $\Omega_A$. This in fact is exactly what happens when the
eccentricity evolution is derived in the situation where the constant
magnitude acceleration rotates at a constant rate.  It turns out that the
corresponding secular problem is also integrable. The eccentricity and
inclination evolution are given by \cite{c10-1}:
\begin{eqnarray}
e^2&=&\frac{p^2\sin^2\alpha}{4 \nu_{+}^2\nu_{-}^2}\left[2  (3 + p^2)
-4 (1+p \cos\alpha  )\ \cos \nu_{+}t 
-4 ( 1- p \cos\alpha)\  \cos \nu_{-}t\right.\nonumber\\
&& +( 1-p^2+\nu_{+}\nu_{-}) \cos (\nu_{+}-\nu_{-})t
\left.+(1-p^2-\nu_{+}\nu_{-})\cos (\nu_{+}+\nu_{-})t\right]\label{e-jet}\\
\cos I&=&\frac{1}{2
  \nu_{+}^2\nu_{-}^2\sqrt{1-e^2}}\left[(p^4-p^2+2+p^2[p^2-3]\cos 2\alpha)\nonumber\right.\\
&&  \ \ \ \ \ \ \ \ \ \ \ \ \ \ +p^2\sin\alpha^2(p^2+1+2 p\cos\alpha)\, \cos\nu_{+}t
\\&& \ \ \ \ \ \ \ \ \ \ \ \ \ \ \left. +p^2\sin\alpha^2(p^2+1-2 p\cos\alpha)\, \cos\nu_{-}t \right]
 \label{i-jet}
\end{eqnarray}
where $\alpha$ is the jet angle with respect to the $z$--axis of the reference
frame, $\nu_{\pm}^2=p^2+1\mp 2 p\cos\alpha$, $p=n_A/2\Omega_A$, and the time
$t$ is normalized by $\Omega_A$. The companion's initial orbit is circular and
lies in the $xy$--plane.
%%%%%%%%%%
%%% FIG  5

\begin{figure}
\centering
\includegraphics[height=3.7cm]{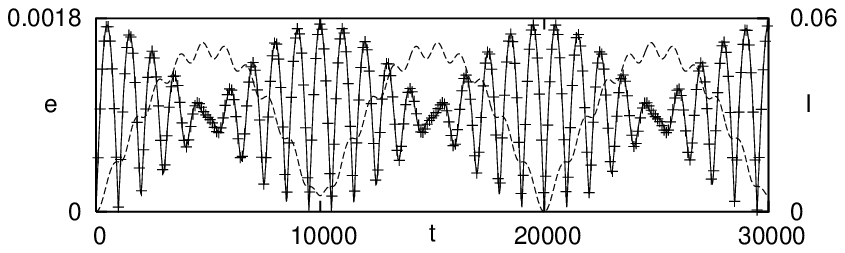}\\
\includegraphics[height=3.45cm]{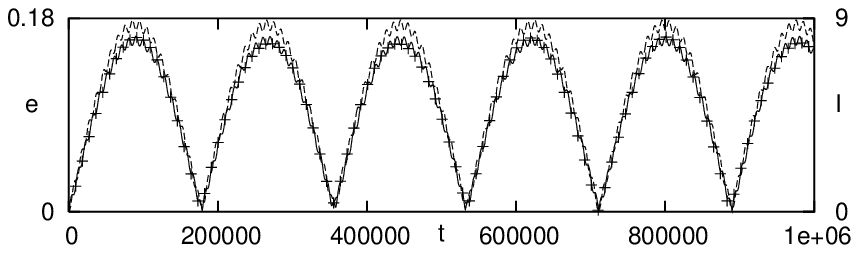}\\
\includegraphics[height=3.3cm]{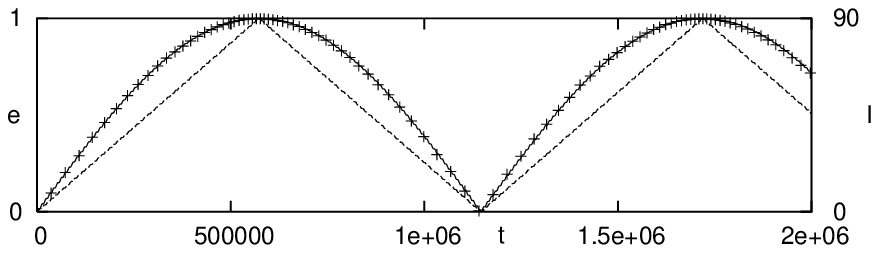}
\caption{Excitation of the eccentricity (solid) and inclination (dashed) by a
  nearly perpendicular precessing jet with an angle $\alpha=1^\circ$.  The
  companion's orbit is located at $a=1$ AU and evolves from a circular orbit
  in a plane orthogonal to the jet's precession axis. The acceleration is
  $A_0=2\times 10^{-10}{\rm km\,s}^{-2}$ yielding an excitation time of
  $2\pi/n_A=10^4$ years. From top to bottom, the panels were obtained from
  equations (\ref{e-jet}--\ref{i-jet}) with the frequency ratios, $p$: 0.05,
  0.9,and 1 -- the precession period is $2p\times 10^4$ years. The symbols
  correspond to the numerical integration of the full equations of motion.}
\label{fig5}
\end{figure}

%%%%%%%%%%
Nominal resonance is defined where the frequency match, $p=1$, occurs. It
corresponds to a nominal resonant semi-major axis $a_{\rm
  res}$ given as:
\begin{equation}
a_{\rm res}\simeq 4 \ \
 \left(\frac{M+m}{M_\odot}\right)\,\left(\frac{A}{2\times 10^{-10}\ {\rm km}\, {\rm s}^{-2}}\right)^{-2}\left(\frac{T_{\rm
 prec}}{10^4\,\mbox{years}}\right)^{-2}\ \ \mbox{AU}, \label{ares}
\end{equation}
where $T_{\rm prec}=2\pi/\Omega_A$. In terms of the resonant semi-major axis,
the frequency ratio can be written as $p=\sqrt{a/a_{\rm res}}$. Far inside
resonance ($p\ll1$), the jet precesses faster than the eccentricity excitation
leading to a reduction of the eccentricity amplitude from $\sin\alpha$ to
$2p\sin\alpha$. Far outside resonance ($p\gg 1$), the jet's precession is slow
compared to the eccentricity excitation so that the latter is described by a
constant acceleration without rotation. In the resonance region, the proximity
of $p$ to unity increases the denominators of the eccentricity expression
(\ref{e-jet}) which leads to eccentricities close to unity. At exact
resonance, the eccentricity reaches unity regardless of the jet angle. The
width of the region around resonance increases with the jet angle $\alpha$.
These features are illustrated in Figures (\ref{fig5}) where we plot the
expressions (\ref{e-jet}) and (\ref{i-jet}) for a jet angle $\alpha=1^\circ$,
an excitation time $2\pi/n_A=10^4$ years, and the three values of $p$: 0.05,
0.9, and 1.  Finally, we note that as the eccentricity excitation time is
$n_A$, no resonant forcing occurs when $\Omega_A=n$ in the secular region
($n_A\ll n$).

\newpage

\subsection{Sudden jet-induced excitation and radial migration}
\begin{figure}\centering
\includegraphics[height=3.7cm]{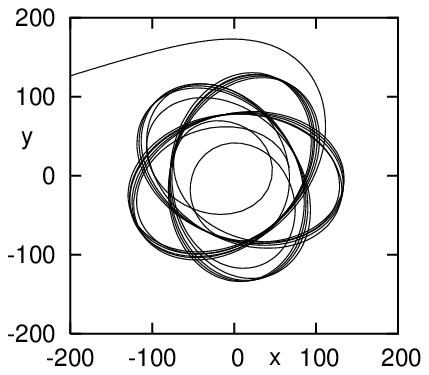}
\includegraphics[height=3.7cm]{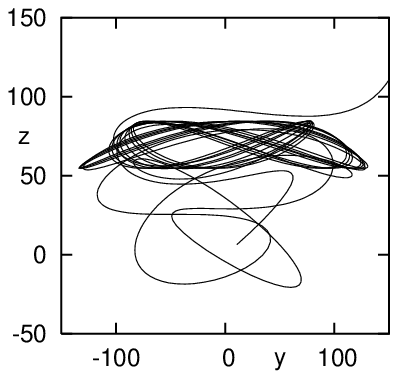}
\includegraphics[height=3.7cm]{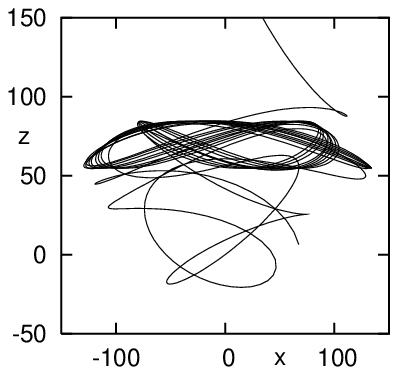}
\caption{Escape of a companion located near the keplerian boundary of a
         conservative constant-direction acceleration with $I_0=30^\circ$. The
         distances are given in AU. The boundary's semi-major axis is $a_{\rm
         kplr}=100$\,AU. Note how the companion hovers above the star before
         escaping.  }
\label{fig6}
\end{figure}

The location where orbits can not longer be retained by the star is where the
frequency $n_A$ becomes comparable to the local mean motion $n$ of the
companion.  Near this limit, the forced periodic oscillations of the
semi-major axis $a$ are reinforced by the eccentricity and acquire large
amplitudes making the orbits unstable in the long term. Calling $a_{\rm kplr}$
the semi-major axis of the keplerian boundary where the star where $n_A=n$, we
the jet-induced acceleration is expressed as:
\begin{equation}
|A_0|\simeq 
2\times 10^{-12}\left(\frac{M+m}{M_\odot}\right)\,\left(\frac{10^3\, {\rm AU}}{a_{\rm
 kplr}}\right)^2\ {\rm km}\, {\rm s}^{-2} \label{stability1}
\end{equation}
corresponding to an excitation period $T_A=2\pi/n_A$:
\begin{equation}
T_A\simeq
10^{6}\left(\frac{M+m}{M_\odot}\right)^\frac{1}{2}\,
      \left(\frac{a_{\rm kplr}}{10^3\, {\rm AU}}\right)^2\,
      \left(\frac{1\, {\rm AU}}{a}\right)^\frac{1}{2}\
   {\rm years} \label{TA}.
\end{equation}
Figure (\ref{fig6}) shows an example of an escape orbit of a
constant-direction acceleration with $a_{\rm kplr}=10^2$\,AU and an
inclination $I_0=30^\circ$. The orbit's initial semi-major axis is 68.5 AU.
The characteristics of motion are not strictly keplerian as the companion
hovers above the star. Such escape orbits offer an interesting way to expel
planets from around their parent stars or equivalently to disrupt a binary
stellar system. If a companion is formed near the keplerian boundary or is
pushed out to it by a possibly remaining inner disk that followed
photo-evaporation \cite{b6}, it could become unbound. 
%%%%%%%%%%
%%% FIG  6

%%%%%%%%%%%
For a realistic jet, the induced acceleration has a finite duration.
Accordingly, $a_{\rm kplr}$ varies in time from infinity before the jet's
launch to a location determined by the strongest acceleration the jet can
provide. Ultimately, the keplerian boundary is pushed out to infinity.  The
acceleration's finite duration extends the keplerian boundary depending on the
ratio of the duration $\tau$ to the excitation time at the keplerian boundary
of the equivalent constant-acceleration problem $T_A(a_{\rm kplr})/2$.  When
$\tau\geq T_A(a_{\rm kplr})/2$, orbits beyond $a_{\rm kplr}$ have enough time
to acquire sufficient momentum and escape the gravitational pull of the star.
When $\tau\leq T_A(a_{\rm kplr})/2$, the stability region extends beyond
$a_{\rm kplr}$. The new stability boundary is given by the semi-major axis
where $\tau\simeq T_A(a_\infty)/2$ which is larger than $a_{\rm kplr}$ since
$T_A$ is a decreasing function of the semi-major axis $a$. The expressions of
$T_A$ and the residual velocity $V$, show that $a_\infty\simeq G(M+m)V^{-2}$,
the location where the keplerian velocity $v$ matches $V$.

Orbits near the keplerian boundary of a finite duration acceleration that do
not escape the pull of the star will end up with elongated orbits whose
semi-major axes and eccentricity have changed. This happens because the
companion feels an almost instantaneous velocity kick (the orbital period is
large compared to $\tau$). The conservation of linear momentum and energy can
be combined to find the change in semi-major axis as:
\begin{equation}
\frac{1}{a_{\rm f}}=\frac{1}{a_{\rm i}}-\frac{2 V\sin
  I_0\cos\theta}{\sqrt{G(M+m) a_{\rm i}}}  -\frac{V^2}{G(M+m)},
\label{sudden3}
\end{equation}
where $a_{\rm i}$, $a_{\rm f}$ are the initial and final semi-major axis,
$\theta$ is the longitude of the companion along its orbit, and $I_0$ is the
inclination of the orbital plane with respect to the direction of the residual
velocity ${\bf V}$. Note that for $I_0\neq 0$, the final semi-major axis can
be larger or smaller than the initial value depending on the longitude of the
companion where the velocity pulse if felt.

%%%%%%%%%%%%
%%%  FIG 7

\begin{figure}
  \centering
  \includegraphics[height=4.3cm]{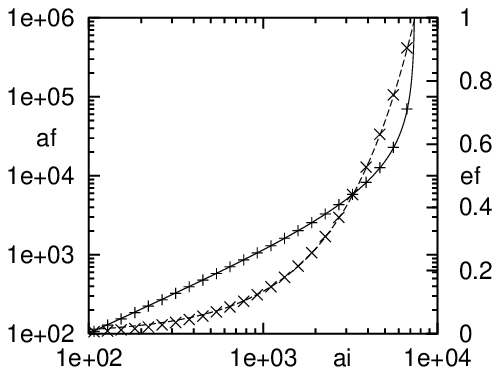}\includegraphics[height=4.3cm]{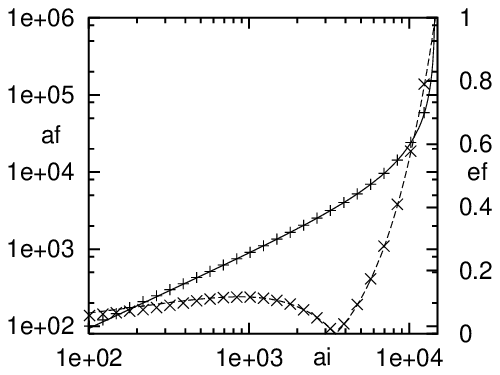}
\caption{Migration and eccentricity excitation near the keplerian boundary of
         a finite duration acceleration for two inclination values $I_0=0$
         (left panel) and $I_0=20^\circ$ (right panel).  In each panel, the
         final semi-major axis $a_{\rm f}$\,(AU) (solid) and the final
         eccentricity $e_{\rm f}$ (dashed) are shown as a function of the
         initial semi-major axis $a_{\rm i}$\,(AU). The parameters are:
         $V=0.35$\,km\,s$^{-1}$, $a_{\rm kplr}=300$\,AU, and $\tau=500$
         years. The inclined circular orbits were started at the descending
         node ($\theta=180^\circ$).  For $I_0=0^\circ$,
         $a_{\infty}=7\,341$\,AU and for $I_0=20^\circ$, 
         $a_{\infty}=14\,542$\,AU.}
\label{fig7}
\end{figure}

%%%%%%%%%%%%%
Figure (\ref{fig7}) shows the final semi-major axes and eccentricities at two
different inclinations $I_0=0$ and 20$^\circ$ for an acceleration amplitude
corresponding to $a_{\rm kplr}=300$\,AU and a duration $\tau=500$ years
resulting in a residual velocity $V=0.35$\,km\,s$^{-1}$. The companions were
started at the descending node for $I_0=20^\circ$ to illustrate inward and
outward migration. Such migration could enhance the delivery of minor bodies
to the Oort Cloud and explain the transport of Kuiper Belt outliers 2000 CR105
and Sedna (90377).

\subsection{A test case: the $\upsilon$~Andromedae binary system}
Multiplanet systems provide good test cases for the excitation by jet-induced
acceleration. This is because mutual gravitational interactions cause the
eccentricity vectors to precess. If the ensuing precession rates are much
faster than the excitation frequencies, the slow build up of eccentricity by
jet-induced acceleration will be lost \cite{c10-1,c10-2}. This situation is
similar to that encountered for precessing jets inside the resonance radius
$a_{\rm res}$.

An interesting system for testing the excitation mechanism is that of
$\upsilon$~Andromedae \cite{b41}. It contains three planets, two of which have
their apsidal directions aligned \cite{b37,b38,b39} as well as a $0.2M_\odot$
stellar companion at a projected distance of 750\,AU \cite{b36}.

Numerical simulation can be used to reproduce the planetary orbits from
initially circular co-planar orbits to their observed state \cite{b40}:
$a_b=0.059$\,AU, $e_b=0.020$, $\omega_b=241^\circ$, $m_b \sin i=0.75\,m_{\rm
  J}$, $a_c=0.821$\,AU, $e_c=0.185$, $\omega_c=214^\circ$, $m_c\sin
i=2.25\,m_{\rm J}$, $a_d=2.57$\,AU, $e_d=0.269$, $\omega_d=247^\circ$, and
$m_d\sin i=2.57\,m_{\rm J}$. It turns out that mutual planetary perturbations
are strong enough to prevent excitation if the acceleration is smaller than
$A_0\sim 10^{-11}{\rm km\,s}^{-2}$. The equivalent smallest keplerian boundary
is at $a_{\rm kplr}\sim 500\,$AU.  Below this value, the current configuration
can be recovered along with the apsidal alignment of the outer two planets.
The presence of the stellar companion outside the keplerian boundary leaves us
two options: either the excitation by acceleration is ruled out or that the
companion was initially inside the boundary and migrated during a sudden
excitation (the projected distance of 750\,AU does not translate necessarily
into a semi-major axis as the companion's orbit is likely to be eccentric).

%%%%%%%%%%%
%%%  FIG 8

\begin{figure}\centering
\includegraphics[height=3.3cm]{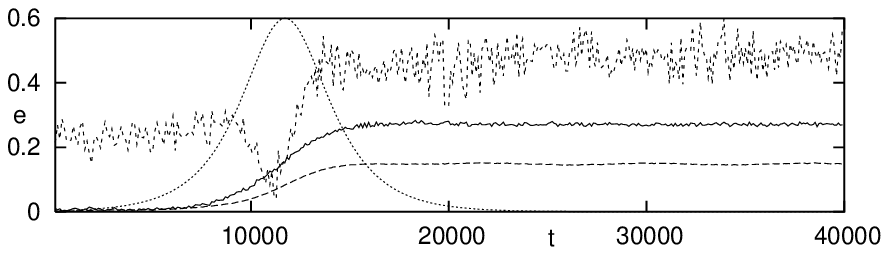}\\
\includegraphics[height=3.3cm]{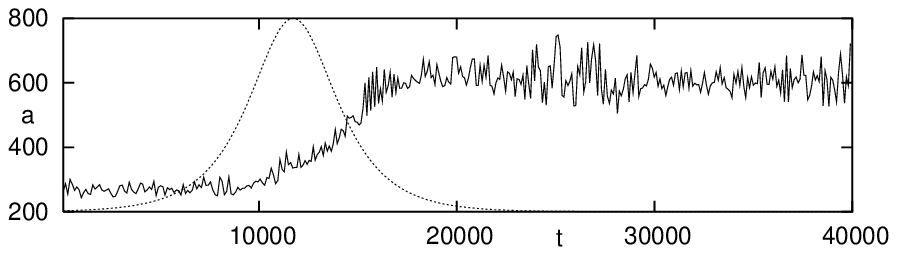}\\
\includegraphics[height=3.3cm]{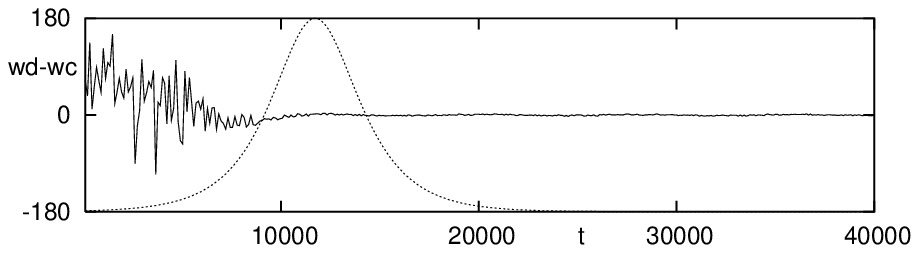}
\caption{Eccentricity excitation,  apsidal alignment and binary migration 
  in the $\upsilon$~And system. The acceleration pulse is shown not to scale in
  all panels.  The first panel shows the eccentricity excitation of planets
  $d$ (solid) and $c$ (dashed) and the eccentricity evolution of
  $\upsilon$~And B (short-dashed) as well as the acceleration pulse normalized
  to its maximum value (dotted).  The second panel shows the radial migration
  of $\upsilon$~And B. The last panel shows the relative apsidal libration of
  planets $d$ and $c$.}
\label{fig8}
\end{figure}

%%%%%%%%%%%

Figure \ref{fig8} shows a simulation of the jet-induced acceleration of the
form $A(t)=A_0/\cosh(t-t_0)/\tau$ where $A_0\sim 3\times
10^{-11}$\,km\,s$^{-2}$ and $\tau=2000$\,years, applied the current planetary
system plus a stellar companion on an orbit of semi-major axis $a=298$\,AU,
just inside the keplerian boundary of $A$, $a_{\rm kplr}\sim 300\,$AU.  The
stellar companion's initial orbit was given an eccentricity $e=0.3$ in order
to decouple its perturbations from the planetary system. In particular, the
eccentricity excitation by the Kozai mechanism \cite{b8-0} is not efficient
because the corresponding excitation time ($\sim 10^7$\,years) is much larger
the duration of acceleration and the eccentricity secular frequency of the
isolated two-planet system ($\sim 7000$\, years). The jet-induced acceleration
produces a configuration similar to the observed one with stellar orbital
elements: $e=0.5$, $a=600$\,AU. Apsidal alignment is achieved as the result of
the acceleration's strength that maintain the forcing of companion orbits to be
perpendicular to the direction of acceleration. Note that only when the
acceleration's strength is near maximum and the keplerian boundary nears
300\,AU, does the stellar orbit acquire a larger eccentricity.

\subsection{The solar system}
Was the solar system subject to jet-induced acceleration?  There are two
observations that hint at the dynamical action of the solar system's jet. The
first is the inclination of Jupiter's orbital normal by 6 degrees with respect
to the sun's rotation axis. As Jupiter is the more massive planet in the solar
system, this implies that either the early protoplanetary disk of the solar
system was warped with respect to the Sun's equator plane or that the
Jupiter's orbit gained inclination with respect to an early equatorial disk.
Both possibilities are consistent with the jet-induced acceleration theory.
More recently, the discovery of Calcium-Aluminum inclusions in the grains of
comet Wild 2 suggest that the only possibility for the Jupiter-family comet
originating from the early Kuiper belt to contain such high temperature
minerals is that they were transported by the solar system's jet. An
additional piece of the puzzle is given by outer solar system bodies such as
Sedna (90377) that are decoupled from the solar system's planets as their
perihelia are larger than Neptune's orbital radius. Such objects could have
been transported by the radial migration in the sudden excitation region
associated with the jet-induced acceleration.

Why are the solar system's planets eccentricities small? There are three
possible reasons for this: first, the resonance radius (\ref{ares}) where the
jet's precession matches the excitation time could have been outside the
planets' orbits. Inside the resonance radius, very little excitation can take
place (Figure \ref{fig5}). Second, mutual planetary perturbations could have
destroyed the secular eccentricity growth. Third, if the jet angle were to be
small and the acceleration duration equally brief, the planets' location near
the resonant radius would be of little help to raise their eccentricities.

\subsection{The unknowns of jet-induced excitation}
The unknowns of jet-induced excitation belong to two categories: (1) the
unknowns of disk-planet interaction and (2) the unknowns of the time evolution
of jets. In the first category comes the issue of the eccentricity damping by
the accretion disk. As explained in the section about mean motion resonances,
the damping of eccentricity for large eccentricity and inclination orbits is
not currently understood. The advantage of jet-induced excitation with respect
to the excitation during migration while in mean motion resonance is that
excitation times can be much smaller than the migration timescale and the
viscous timescale of the disk. In this phase, the formed planets gain a
substantial inclination that makes them exit the gas disk which should in
principle reduce the eccentricity damping significantly. In this context, it
is useful to bear in mind that jet-induced acceleration becomes effective in a
planetesimal disk only when a few planets are left. The planetesimal mutual
gravitational interactions destroy excitation through the radom precession of
their orbits. When the bodies left in the disk are such that the precession
periods due to their mutual perturbations are larger than the excitation time,
jet-induced acceleration becomes effective and planets may exit the disk on
inclined orbits. For the second category of unknowns, it is safe to say that
except for precession, little information is available about the time
variations of jets over their entire life span.  Perhaps the main advantage of
jet-induced excitation is its small set of parameters: amplitude, duration and
jet precession frequency.  These determine all the features of eccentricity
growth or lack thereof. In multiplanet systems, the acceleration subjects all
companions, planetary and stellar alike, to the same instability as it is
independent of position and velocity. A better knowledge of the time
dependence of acceleration can therefore easily confirm or rule out the effect
of jet-induced excitation in multiplanet systems.

\section{Concluding remarks}

The various theories of eccentricity excitation are valuable tools to gain
insight into one of the most pressing if not the most pressing issue in planet
formation theory: the mismatch of protoplanetary disk lifetimes and the
timescales of planet formation and planet migration. Protoplanetary disk
lifetimes range from $10^5$ to $10^7$ years. To have a viable theory, the
planet formation timescale has to be smaller by at least an order of magnitude
than these estimates. Planets are believed to form in two ways: (1) rocky core
accretion and (2) gravitational instability. The formation of rocky cores
precedes gas accumulation on the way to forming gaseous planets. This specific
phase is slow. Typical formation timescales range from $10^6$ to $10^7$ years
at Jupiter's location \cite{d2}.  Gravitational instability however has a much
smaller timescale of order $10^3$ years at the same location. With all its
caveats about the required disk opacity for gravitational fragmentation and
the role of shearing instabilities in disrupting a forming protoplanet,
formation by gravitational instability looked far more promising than rocky
core accretion to explain the existence of the hot Jupiters. This was true
until the discovery of a hot Saturn with a 70 Earth mass rocky core orbiting
the star HD 149026 \cite{d3}--for reference, the mass of Jupiter's rocky core
is believed to be 15 Earth masses. The formation of a 70 Earth mass rocky core
either as a single body or as a merger of smaller cores is likely to last at
least $10^7$ years according to current planetesimal accumulation models. This
temporal crisis does not stop at rocky core formation, it worsens because of
planet migration. Migration arises from the planet's interaction with the gas
disk and has typical timescales of $10^6$ years for an Earth mass core (linear
regime, type I) to $10^5$ years for a Jupiter mass planet (nonlinear regime,
type II) \cite{b25,d1}. The duration of the eccentricity excitation phase and
its dependence on the mechanism's parameters may help elucidate the time
sequence of the events that produced the observed extrasolar planets. In
particular, three-dimensional hydrodynamic simulations of disk-planet
interactions for eccentric and inclined orbits (off the disk's mid-plane) may
elucidate the problem of the end of resonant excitation in multiple systems.
Planet-planet scattering models may benefit from including a wider mass
spectrum in the populations of simulated planets.  The excitation by Kozai's
secular mechanism requires further assessments of the ability to form multiple
planets under the stellar companion's perturbations \cite{b23-1}. Similarly,
jet-induced excitation was demonstrated for a single planet under the action
of a precessing acceleration. The effect of mutual planetary entrainment for
planets on both sides of the precession--excitation resonance needs to be
investigated and applied to observed extrasolar systems.

\end{document}